\documentclass[sigconf]{acmart}
\AtBeginDocument{%
  \providecommand\BibTeX{{%
    \normalfont B\kern-0.5em{\scshape i\kern-0.25em b}\kern-0.8em\TeX}}}
\usepackage{amsmath,amsfonts}
\usepackage{algorithm}
\usepackage{algpseudocode}
\usepackage{graphicx}
\usepackage[caption=false]{subfig}
\usepackage{float}
\usepackage{multirow}
\usepackage{array}
\graphicspath{{figs/}}
\usepackage{hyperref}

\copyrightyear{2022} 
\acmYear{2022} 
\setcopyright{rightsretained} 
\acmConference[SNTA'22]{Proceedings of the Fifth International Workshop on Systems and Network Telemetry and Analytics}{June 30, 2022}{Minneapolis, MN, USA}
\acmBooktitle{Proceedings of the Fifth International Workshop on Systems and Network Telemetry and Analytics (SNTA'22), June 30, 2022, Minneapolis, MN, USA}
\acmDOI{10.1145/3526064.3534111}
\acmISBN{978-1-4503-9315-7/22/06}

\begin{document}
\fancyhead{}
\title{Studying Scientific Data Lifecycle in On-demand Distributed Storage Caches}
%

\author{Julian Bellavita}
\affiliation{%
  \institution{University of California, Berkeley}
  \city{Berkeley}
  \state{CA}
  \country{USA}
}
\email{jbellavita@berkeley.edu}

\author{Alex Sim}
\author{Kesheng Wu}
\affiliation{%
  \institution{Lawrence Berkeley Nat'l Laboratory}
  \city{Berkeley}
  \state{CA}
  \country{USA}
}
\email{{asim,kwu}@lbl.gov}

\author{Inder Monga}
\author{Chin Guok}
\affiliation{%
  \institution{Energy Sciences Network}
  \city{Berkeley}
  \state{CA}
  \country{USA}
}
\email{{imonga,chin}@es.net}

\author{Frank W\"{u}rthwein}
\author{Diego Davila}
\affiliation{%
  \institution{University of California, San Diego}
  \city{La Jolla}
  \state{CA}
  \country{USA}
}
\email{{fkw,didavila}@ucsd.edu}

\renewcommand{\shortauthors}{Author, et al.}

\begin{abstract}
The XRootD system is used to transfer, store, and cache large datasets from high-energy physics (HEP).
In this study we focus on its capability as distributed on-demand storage cache.
Through exploring a large set of daily log files between 2020 and 2021, we seek to understand the data access patterns that might inform future cache design.
Our study begins with a set of summary statistics regarding file read operations, file lifetimes, and file transfers.
We observe that the number of read operations on each file remains nearly constant, while the average size of a read operation grows over time.
Furthermore, files tend to have a consistent length of time during which they remain open and are in use.
Based on this comprehensive study of the cache access statistics, we developed a cache simulator to explore the behavior of caches of different sizes.
Within a certain size range, we find that increasing the XRootD cache size improves the cache hit rate, yielding faster overall file access.
In particular, we find that increase the cache size from 40TB to 56TB could increase the hit rate from 0.62 to 0.89, which is a significant increase in cache effectiveness for modest cost.
\end{abstract}

\ccsdesc[500]{Information systems~Information lifecycle management}
\ccsdesc[500]{Information systems~Hierarchical storage management}
\ccsdesc[500]{Computing methodologies~Model development and analysis}

\keywords{distributed cache, XRootD, data, file, lifecycle}

\maketitle

\section{Introduction}
\label{sec:intro}
Scientific researches are increasingly relying on substantial data for analysis~\cite{Hey:2009:FPD, Evans:2008:LHC}.
In high-energy physics (HEP), a majority of data is stored, transferred, and cached via the XRootD system~\cite{Brun:1997:ROO, xrootd2005}.
Like many scientific research communities, the HEP community collectively has generated a very large volume of data that is widely used by individual researchers around the globe~\cite{osg, xrootdcms}.
To effectively distribute the data to the community, there is a hierarchy of shared sites to replicate the commonly used subsets of data~\cite{xrootdcms, Apollinari:2015:HL-LHC}.
Alongside of this tiered storage system, there is also a collection of distributed data caches to further bring the data closer to the end users~\cite{stashcache, xcache2014}.
This is a study of one of the regional caches to understand the data access patterns and effectiveness of these distributed storage caches.

Distributed storage caches is widely used for large-scale scientific research~\cite{Tierney:1999:DSC},
as well as internet businesses~\cite{onedata, Lannon:2022:Cyberinfrastructure, Syndicate}.
These storage cache systems bring remote data content closer to the users, which reduces the data access time and reduces the demand on the internet backbone.
In scientific research, these distributed caches allow scientists to access large amounts of community data without investing in significant storage resources.
It is an important strategy to democratize large-scale data-intensive scientific research.

As many scientific communities are considering installing such storage caches, it is important to understand how they could be effectively provisioned~\cite{socalrepo2018}.
In this work, we study the usage of currently deployed storage caches in California, USA for the local HEP community.
Based on a study of nearly one-year's history of data access through this cache, we propose a cache simulator to explore resource provisioning options.
More specifically, we investigate the relationship between cache size and cache hit rate.
The currently deployed cache size is 40TB.
Our simulation finds that increasing the cache size to 56TB could increase the cache hit rate from 0.62 to about 0.89.
This is a significant increase in cache effectiveness for a relatively modest cost.

The remainder of this paper is organized as follows.
In Section~\ref{sec:bg}, we provide more detailed background information about the HEP applications and the XRootD software system used for the distributed storage cache system.
This section also provides an overview of the log file used for our work.
In Section~\ref{sec:lifecycle}, we describe the access patterns from the current installation of the storage caches.
This also provides the basis for our cache simulation work in Section~\ref{sec:cachesim}.
A concrete objective of the cache simulator is to explore the resources required for future caches.
We provide a discussion of the statistics and cache simulation results in Section~\ref{sec:discussion}.
We conclude this paper with a brief summary in Section~\ref{sec:summary}.

\section{Background}
\label{sec:bg}

\subsection{High Energy Physics}
The High Energy Physics (HEP) community is among the largest users of global scientific research and engineering networks.
This community depends on unique instruments operated by collaborations across hundreds of institutions around the globe.
Instruments such as ATLAS and CMS at the LHC in Geneva, Switzerland could be thought as high-speed camera with 100 Million pixels capable of capturing many millions of pictures per second~\cite{Evans:2008:LHC}.
With complex real-time decision logic, implemented via a mix of custom hardware and software, these instruments only retain a small fraction of the most interesting data records, known as HEP collision events.
Even after this substantial data reduction, the data volume captured per year is still reaching many petabytes per instrument.
The data volumes are expected to grow by more than an order magnitude by 2028, as a result of detector and collider upgrades for the so-called "High Luminosity LHC" (HL-LHC) science program~\cite{Apollinari:2015:HL-LHC}.


To prepare for this significant increase in data volumes, the LHC community is driven towards making any and all data placement much more dynamic.
A conceptual design under investigation is to replicate data to regional "Data Lakes" and use a mixture of remote access and caching with those lakes~\cite{xrootdcms, datalakes}.
These computing and storage resources are provided by participating countries, typically as in-kind contributions to the collaborations.
In Europe, these "Data Lakes" are less than 1000KM away, where the network latency is low enough that data sharing among the regional institutions is  effective.
In the US, "Data Lakes" are under active investigation to improve application performance. 
This work is a part of this exploration.

\subsection{XRootD System}


The XRootD software suite is a key software in the HEP community and also contains tools for implementing a "Data Lake" as a federated storage cache infrastructure~\cite{xrootd2005, xrootdcms}.
In particular, the StashCache service based on XCache is used by a number of institutes~\cite{stashcache, xcache2014, socalrepo2018}.
It interacts with a tiered data distribution framework, provides storage devices as disk caches, supports distributed data access, and implements a data discovery protocol for dynamic discovering the physical location of objects or files in the logical namespace\cite{socalrepo2018, Fajardo2020}.
Overall, this distributed caching follows a tree-like architecture where each XCache installation could be thought of as a part of the tree~\cite{xcache2014}.
An XCache installation may have a distributed set of servers forming a logical data cache connected to a higher level branch.
Each top level branch of the XCache hierarchy is responsible for a subset of the federated namespace.
Applications are expected to connect to a "regional" cache via the configuration of their runtime environment, e.g. the OSG Data Federation~\cite{Weitzel2017, stashcache, Fajardo2020}.
Cache misses are handled by XCache as XRoot-client calls to the data federation.
Thus, the StashCache service provides relatively low latency data access to the large data collection located far away.
This allows physicists around the word to conduct their data analyses on "small" computer clusters with very limited storage resources, as long as there is a regional XCache nearby.

\subsection{Server Logs and Programming Tools Used}
In this paper we describe the patterns of the data lifecycle observed in one of the XCache nodes at ESnet in Sunnyvale, CA from the Southern California Petabyte Scale Cache~\cite{socalrepo2018} for US CMS which is a part of the Caltech and UCSD Tier-2 center infrastructure. This XCache node has a total cache size of 40TB.
The XCache node, running xrd version v5.1.1, produces daily server logs, which contain information regarding various operations on the cache data. The bulk of the XRootD server logs analyzed in this paper are from the time period between January 2021 and September 2021. The sizes of the daily server logs vary substantially; some are a few hundred megabytes, while others are upwards of 60 Gigabytes. The server logs were processed and analyzed using the NERSC Jupyter system, running on Cori. Using the standard Python library, we parsed each log, searching for keywords or keyphrases denoting specific operations. Lines denoting the desired operation were processed to extract information about the operation, such as the file being operated upon, the size of the operation, and so-forth. We used this extracted information to compute summary statistics regarding file operations and cache behavior. This work provides a comprehensive descriptive analysis of XCache behavior throughout 2021, and hints at changes to the XRootD caching protocol that can improve performance.


\section{File Access Patterns and File Lifecycles in Distributed Caches}
\label{sec:lifecycle}

This section details our findings with respect to file read operations, file lifetimes, and file transfers. 


\subsection{Statistics of File Reads}
\label{subsec:file_stats}
File read operations are denoted in XRootD server logs by two keyphrases: \texttt{fh=0 readV} and \texttt{req=read}. The first keyphrase denotes a readV operation, which extracts a specified number of bytes from a file, beginning at a specified byte offset. The second keyphrase denotes a standard read operation. If a server log line contains either of these keyphrases, then a read operation is being performed during the corresponding timestamp. By parsing the logs while searching for these keywords, it is possible to count the total number of read operations issued in a given time frame. We seek to map these read operations to specific files.

The server logs also indicate the name of the file that each read request is issued towards. For lines that include the phrase \texttt{req=read}, the filename can be extracted directly from the same line of the server log. Lines that include the keyphrase \texttt{fh=0 readV} do not have the filename included, but we can identify the filename by extracting the thread ID and user ID from these lines and matching them with a file open request. File open operations do include the filename, so we can identify the filename corresponding to readV operations by examining their file open operations. Thus, it becomes possible to count the total number of read operations issued to each individual file in the span of the analysis time frame. 

\begin{figure}
    \centering
    \includegraphics[width=0.9\linewidth]{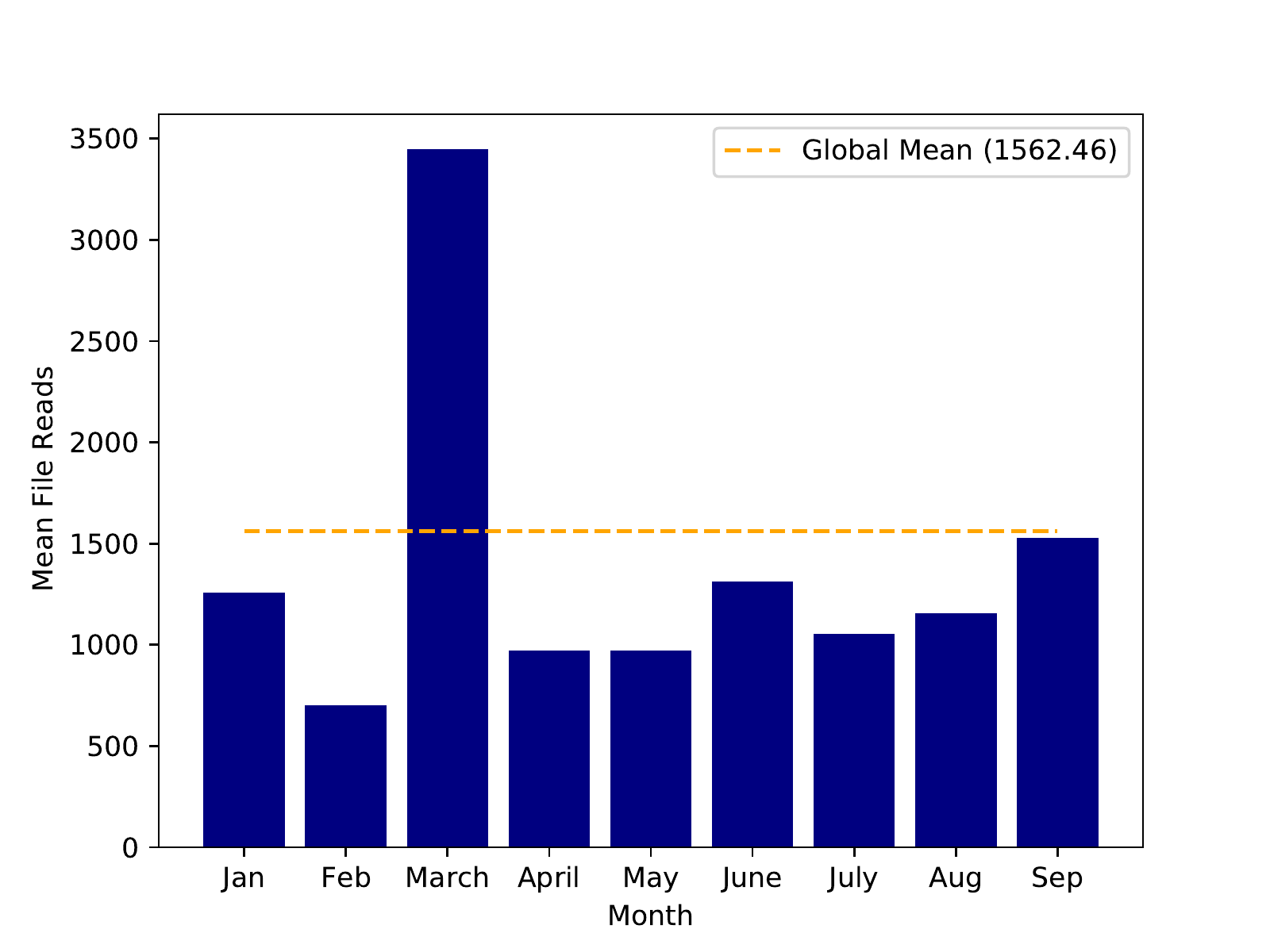}
    \caption{Mean number of read operations issued per-file for Jan 2021--Sep 2021. Global mean = 1562.46} 
    \label{fig:read_rates}
	\vspace{-0.5cm}
\end{figure}

\begin{figure}
    \centering
    \includegraphics[width=0.9\linewidth]{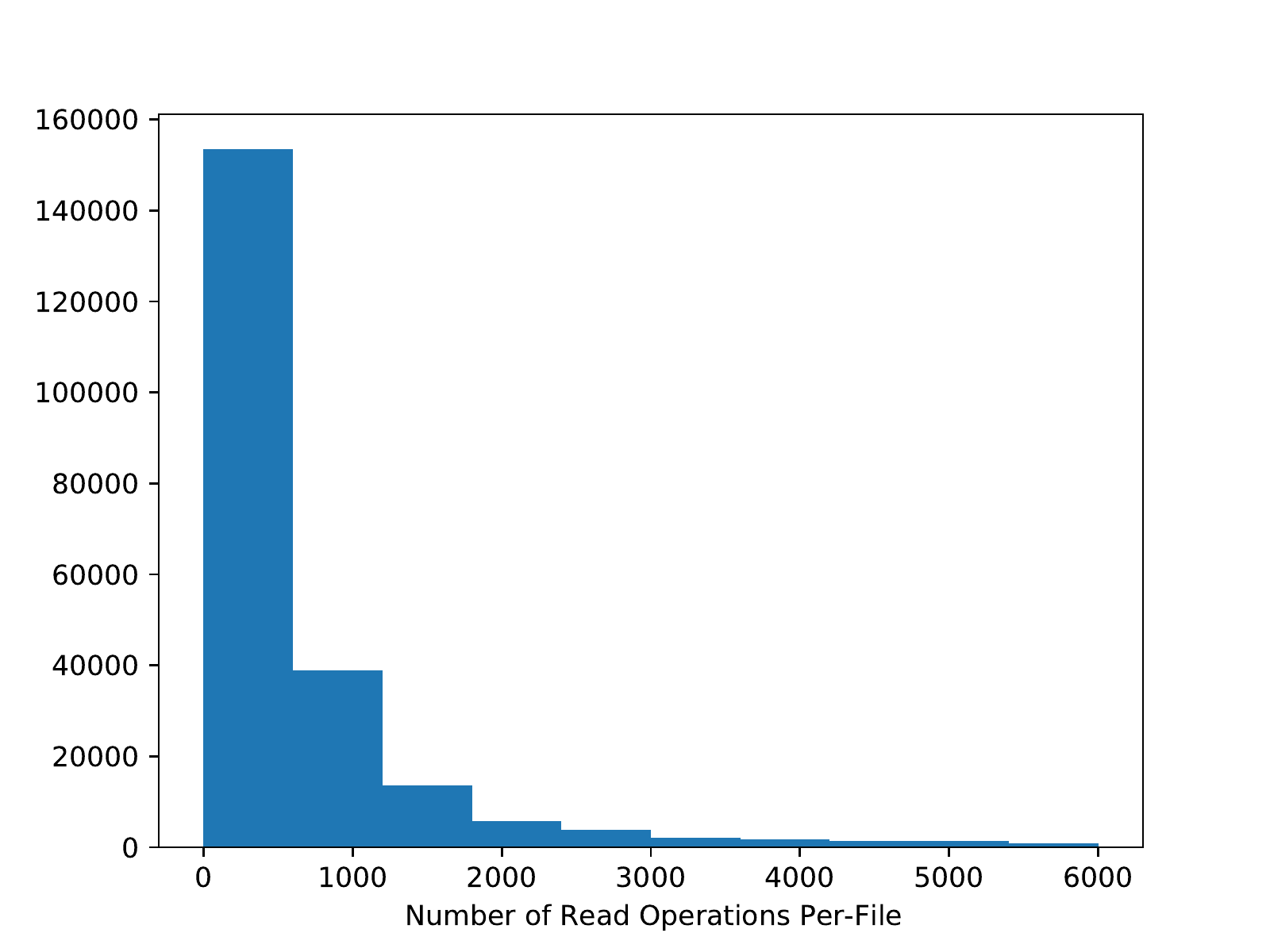}
    \caption{Distribution of monthly read operations per-file for Jan 2021-Sep 2021}
    \label{fig:big_reads}
	\vspace{-0.3cm}
\end{figure}

\begin{figure}
    \centering
    \includegraphics[width=0.9\linewidth]{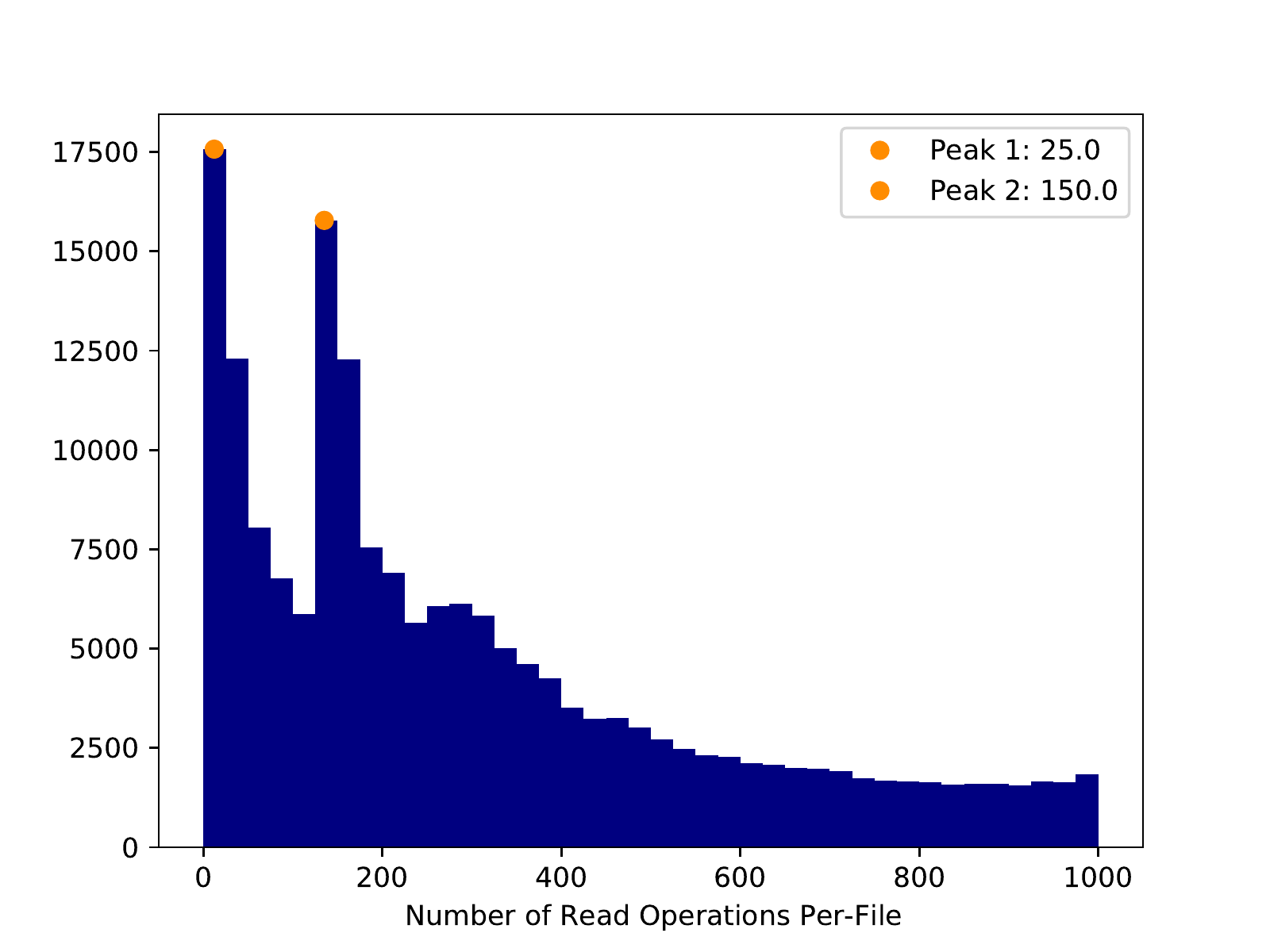}
    \caption{Zoomed-in, finer-grained distribution of total monthly read operations per-file for Jan 2021-Sep 2021. Peaks are at 25 and 150 read operations.}
    \label{fig:reads_fine_grained}
	\vspace{-0.2cm}
\end{figure}

\begin{figure}
    \centering
    \includegraphics[width=0.9\linewidth]{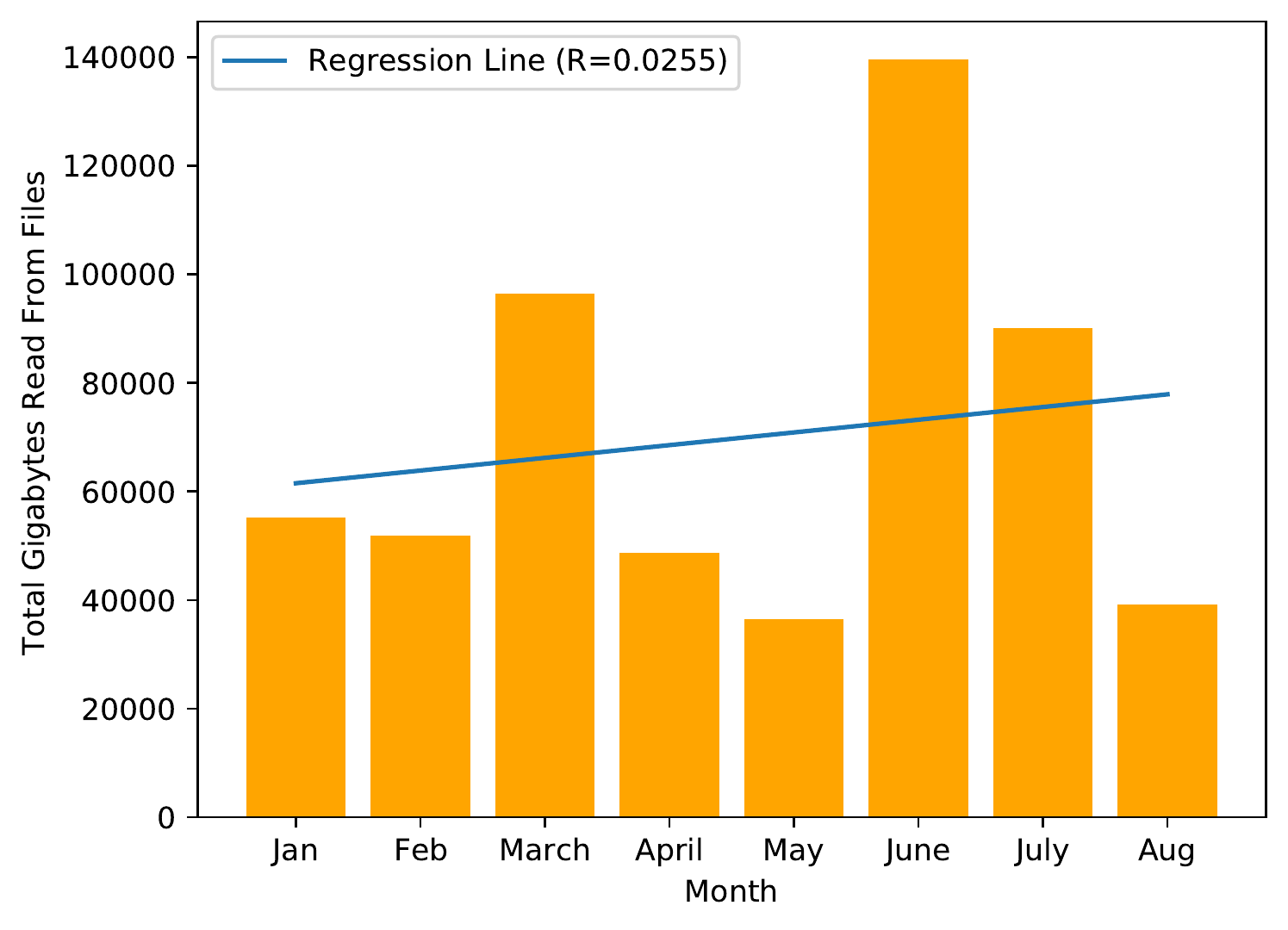}
    \caption{Monthly total size of file reads for Jan 2021-Aug 2021.}
    \label{fig:tot_reads}
\end{figure}

Our procedure for counting file read operations begins by parsing the XRootD logs corresponding to the analysis time frame. It matches readV operations with the appropriate file name using the process outlined above. It then counts the number of read operations issued to each file, using a dictionary to map file names to their read request totals. Once the procedure has finished parsing the server logs, it calculates the mean number of read operations per file by iterating through the dictionary and computing the mean of the set of values in the dictionary. The results of running this procedure for each month in the range January 2021-August 2021 are summarized in Figure \ref{fig:read_rates}. We modified the procedure to allow it to compute and plot the distribution of total monthly read operations per-file, in addition to the mean. Figure \ref{fig:big_reads} depicts the complete range of the distribution, and Figure \ref{fig:reads_fine_grained} depicts a finer-grained view of a subset of the distribution. Additionally, the total number of read operations issued among all files per-month are depicted in Figure \ref{fig:tot_reads}.



\begin{figure}
    \centering
    \includegraphics[width=0.9\linewidth]{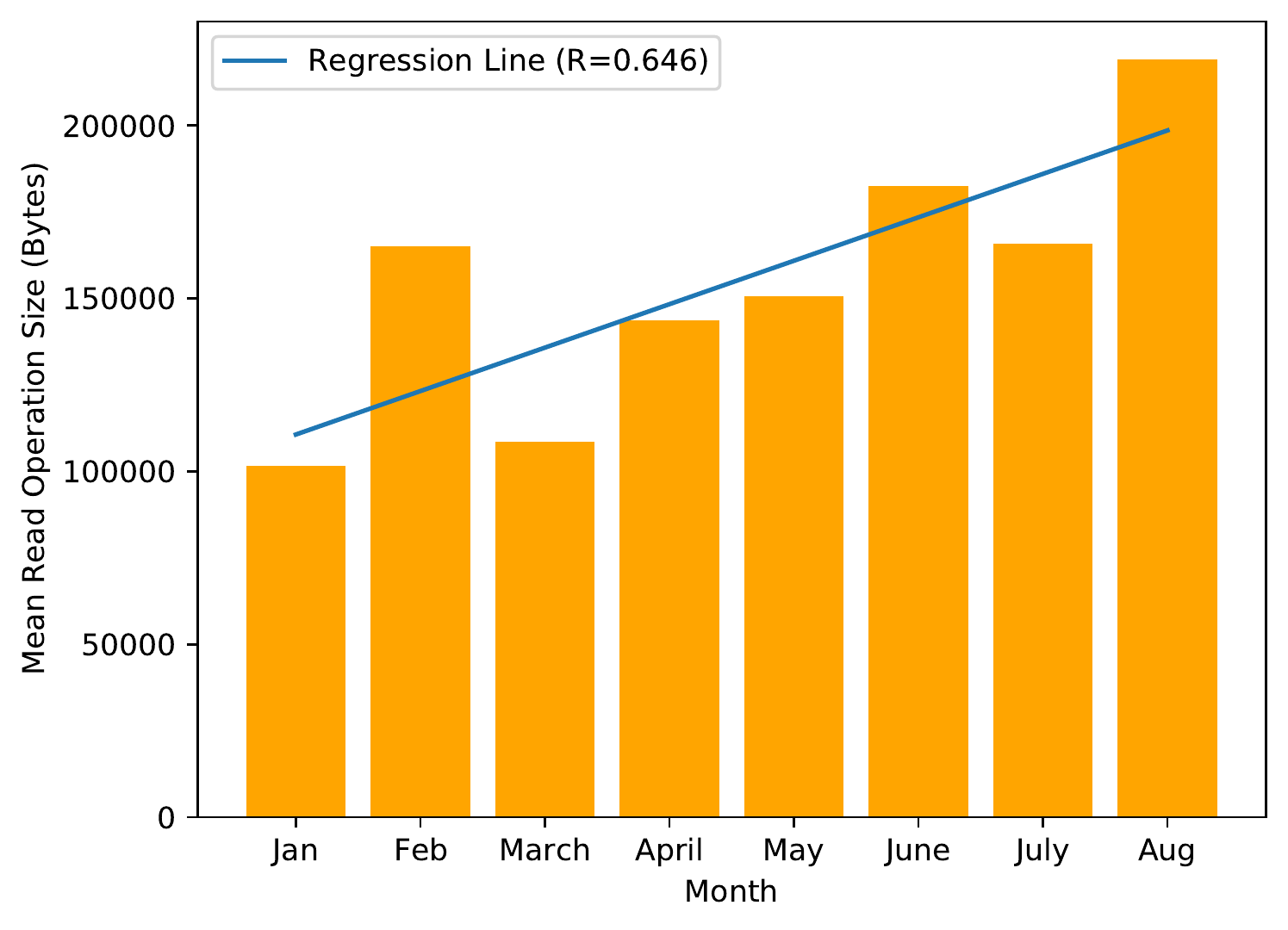}
    \caption{Mean size of file read operations for Jan 2021-Aug 2021. Global mean = 154,632B}  
    \label{fig:mean_reads}
	\vspace{-0.3cm}
\end{figure}

\begin{figure}
    \centering
    \includegraphics[width=0.9\linewidth]{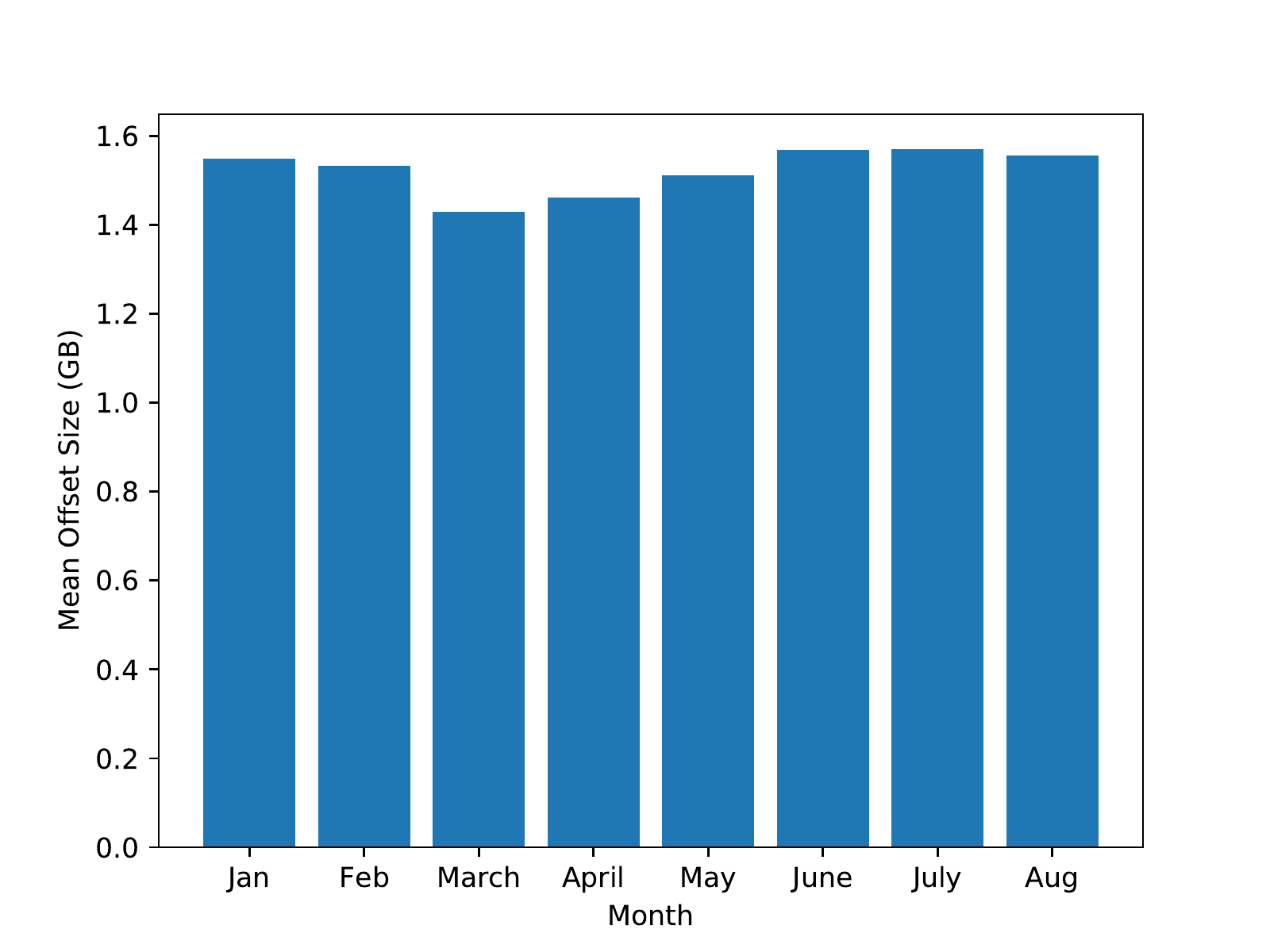}
    \caption{Mean offset size for read operations from Jan 2021-Aug 2021. Global mean = 1.52GB} 
    \label{fig:offsets}
\end{figure}

Read operations and readV operations appear in the server logs in the form \texttt{"NNN@MMM"}, where \texttt{"NNN"} is an integer denoting the number of bytes the user wishes to read from a file, and \texttt{"MMM"} is an integer denoting the byte offset where the user wishes to begin reading bytes from the file. We developed a procedure that parses XRootD server logs, searching for this pattern where it appears in the same line as a read operation or a readV operation. When the procedure locates an instance of this pattern, it extracts the integer representing the size of the read operation, and it extracts the integer representing the offset of the read operation, adding each to a separate running total. This procedure returns the total number of bytes read from files, the mean read operation size, and the mean read operation offset size. The results of running this procedure for XRootD logs spanning Jan 2021 - Aug 2021 are summarized in Figures \ref{fig:tot_reads}, \ref{fig:mean_reads}, and \ref{fig:offsets}. 

\subsection{File Lifetimes}
It is better for files to remain in the XRootD cache only for as long as they remain in use by users. If files stay in the cache for too long, they unnecessarily clog the cache, and if they are evicted from the cache too soon, they must be transferred back from the data sources upon further access. Thus, it is beneficial to know how long files tend to remain open. Knowledge regarding file lifetimes can enable the development and implementation of cache eviction policies superior to the Least Recently Used (LRU) policy. \cite{monjalet2019, luis2021}. 

\begin{figure}
    \centering
    \includegraphics[width=0.9\linewidth]{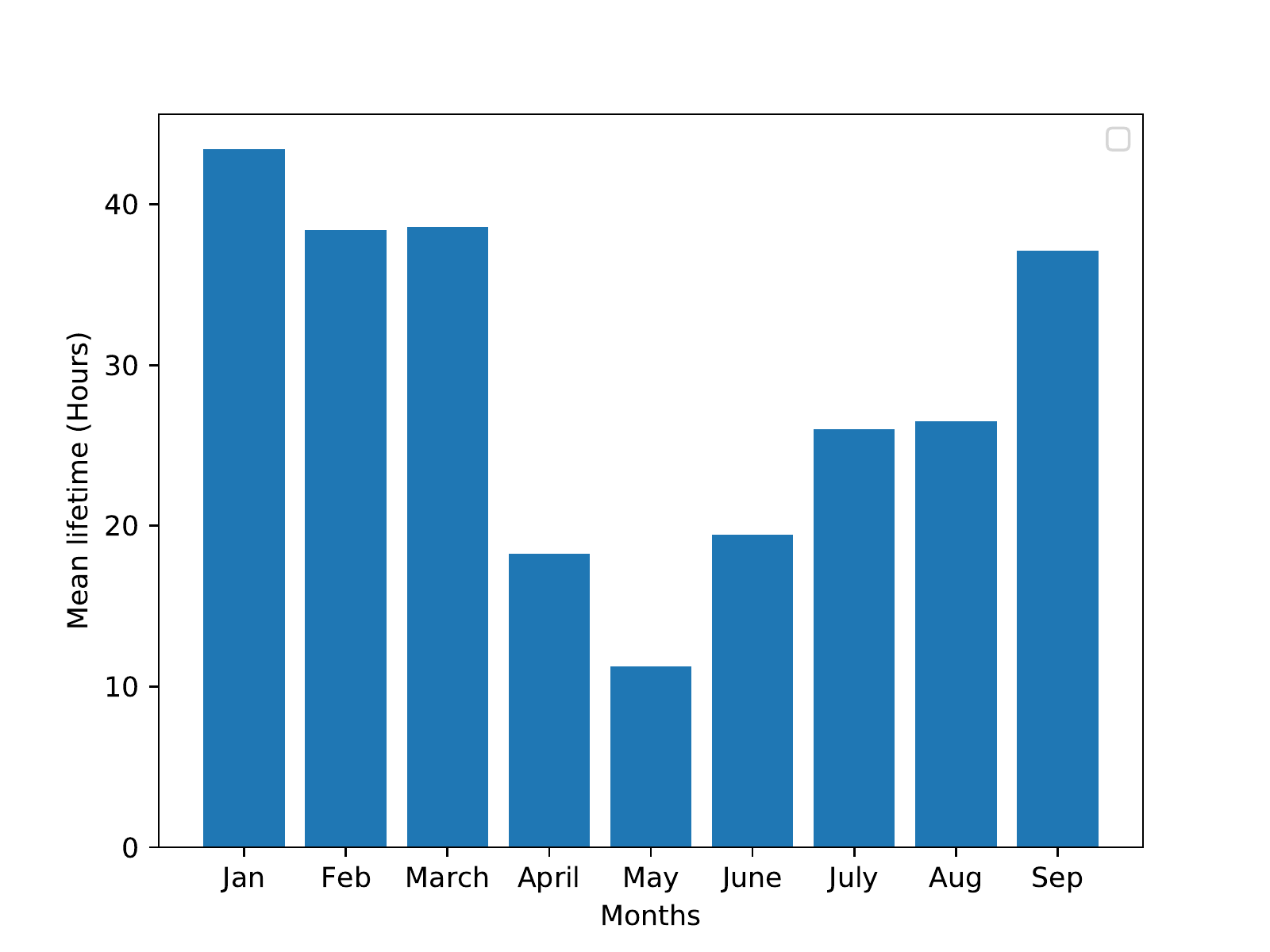}
    \caption{Mean XRootD file lifetimes from Jan 2021-Sep 2021 for threshold points from 1-10 days The global mean of $\overline{X}=28.78$ hours or 1.2 days is used as the threshold for the lifetime experiment.} 
    \label{fig:cutoff}
\end{figure}

To model file lifetimes, we use a standard Python dictionary to map filenames to lists of tuples $(t_s, t_e)_i$, where $t_s$ denotes the timestamp corresponding to the file's first open request in lifetime $i$, and $t_e$ denotes the timestamp corresponding to the file's latest close request in lifetime $i$. Therefore, the length of an arbitrary file lifetime is $t_e-t_s$. A file lifetime is defined as a period of time during which a file is issued an open request at least once every 1.2 days. In other words, if a file goes more than 1.2 days without being opened, its lifetime is considered over. If a file is accessed after the 1.2 day threshold, it is considered the start of a new lifetime. The 1.2 day threshold was computed by modeling file lifetimes across a range of different threshold points (1 day to 10 days), taking the mean lifetime produced by each threshold, and taking the mean of this set of resulting means. Figure \ref{fig:cutoff} summarizes these results.

\begin{figure}
    \centering
    \includegraphics[width=0.9\linewidth]{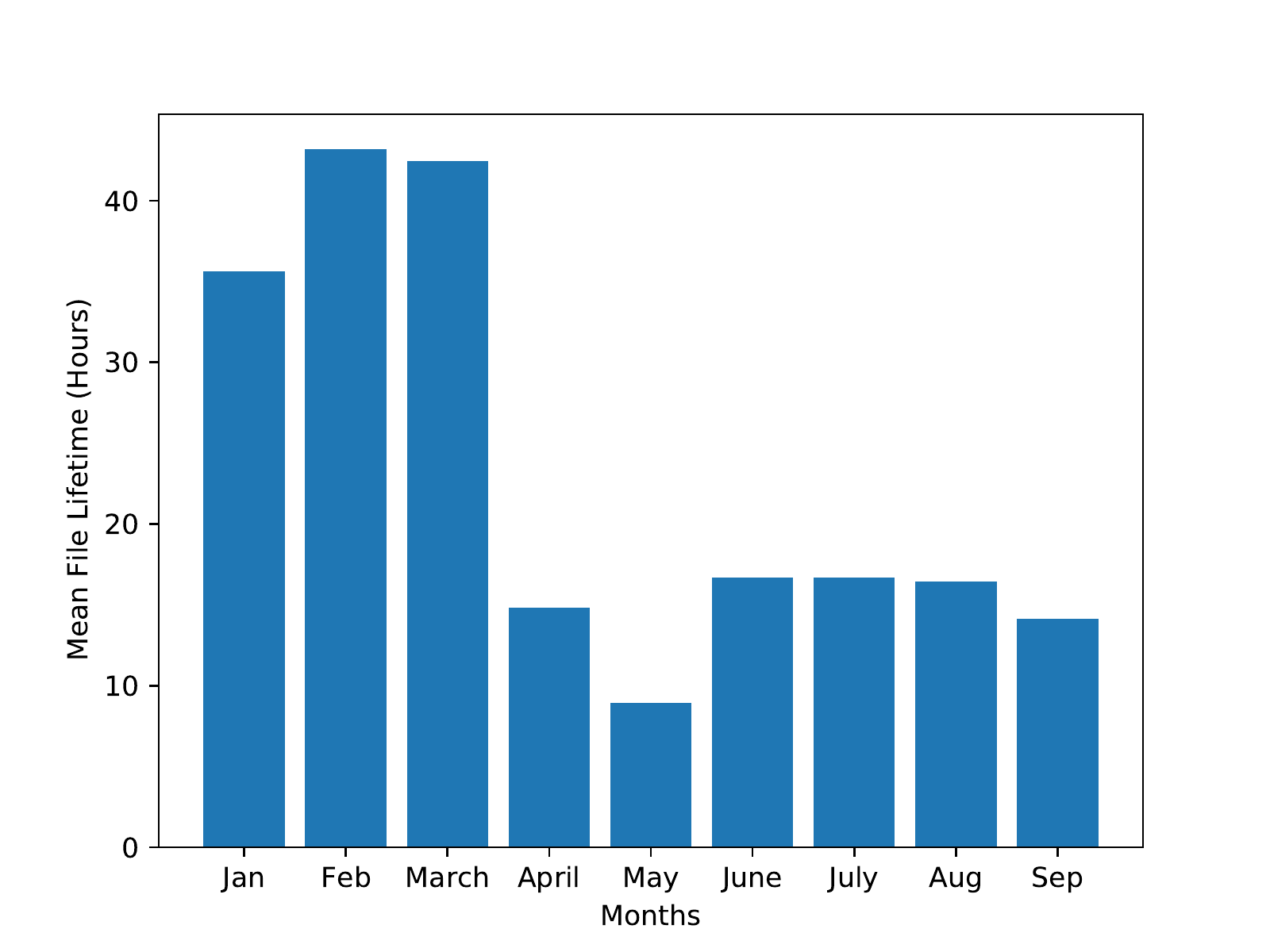}
    \caption{Mean XRootD file lifetimes from Jan 2021-Sep 2021 using a threshold of 1.2 days. Global mean = 23.23 hours. }
    \label{fig:lifetimes}
	\vspace{-0.3cm}
\end{figure}

\begin{figure}
    \centering
    \includegraphics[width=0.9\linewidth]{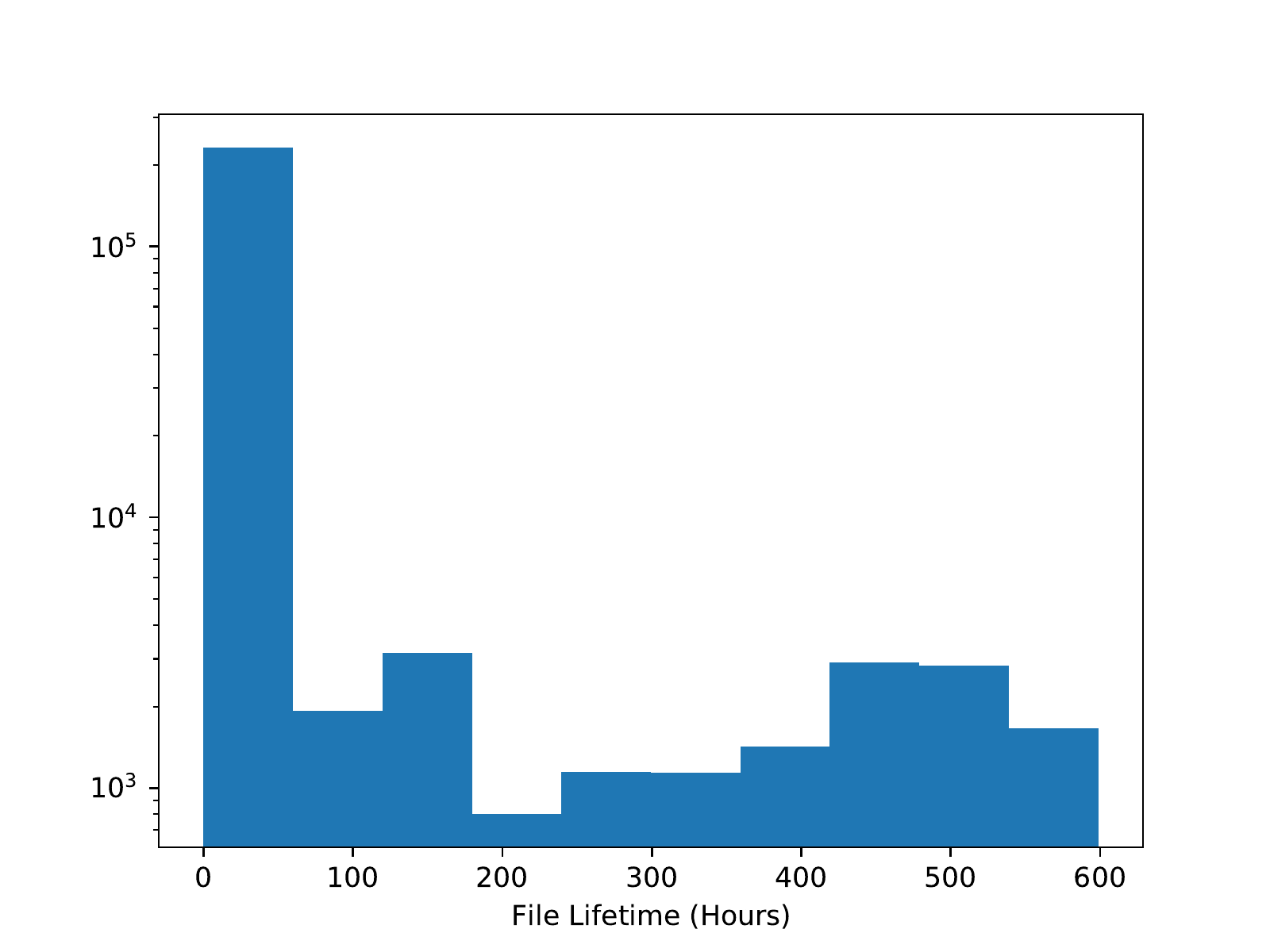}
    \caption{Distribution of file lifetimes for Jan 2021-Sep 2021.}
    \label{fig:lifetime_hist_hours}
\end{figure}

\begin{figure}
    \centering
    \includegraphics[width=0.9\linewidth]{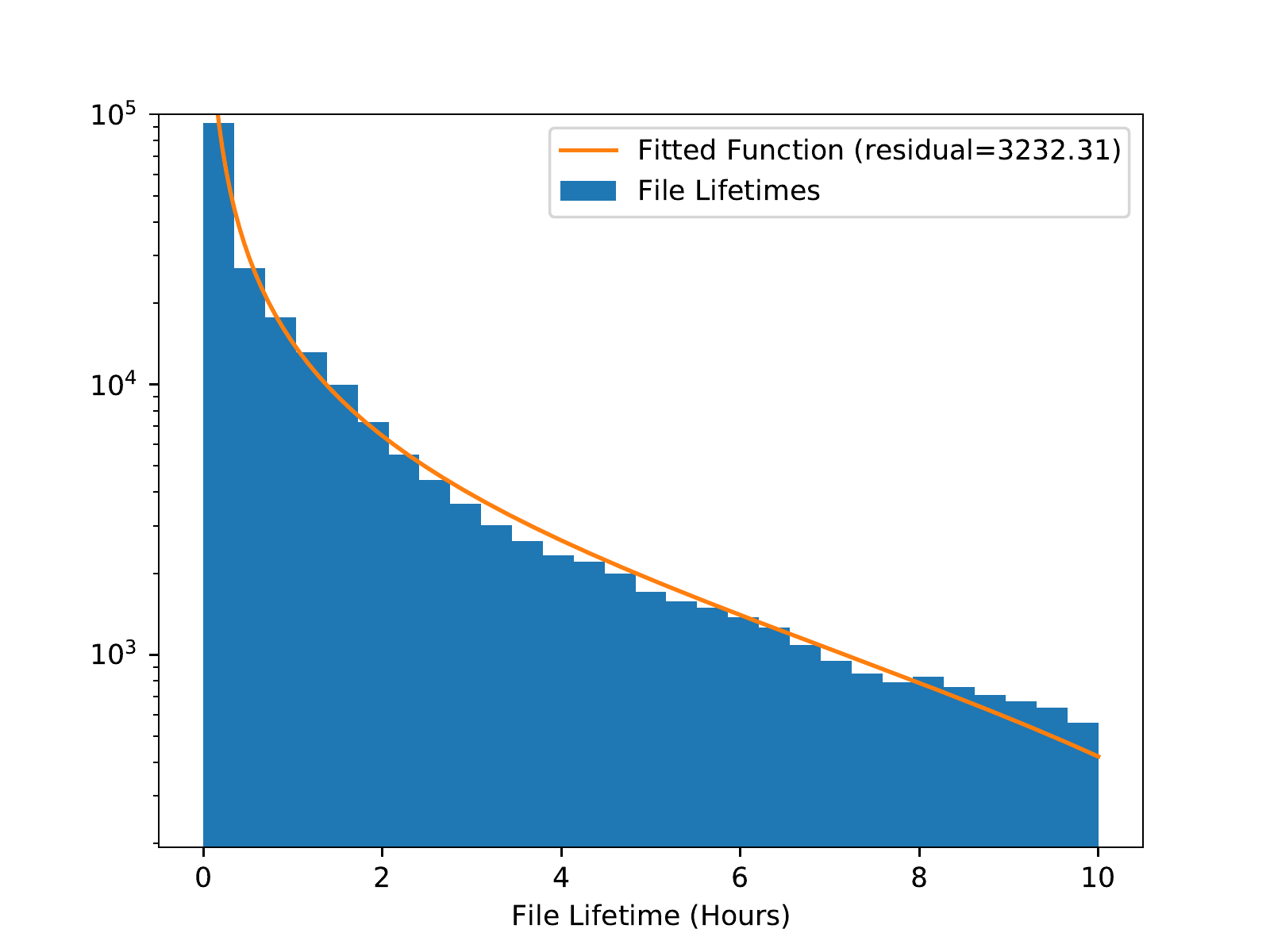}
    \caption{Zoomed-in, finer-grained distribution of file lifetimes for Jan 2021-Sep 2021.}
    \label{fig:medium_lifetimes}
\end{figure}

File open requests are denoted in XRootD server logs by the keyphrase \texttt{"open rat"} or the keyphrase \texttt{"open r"}. File close requests are denoted by the keyphrase \texttt{"prefetch score"}. Server log lines that contain these keyphrases also include the timestamp of the operation and the file name. Thus, these lines provide all the information needed to populate our dictionary. Our procedure parses XRootD server logs while searching for these keyphrases, populates the dictionary with key-value pairs of the form outlined above, and iterates through these pairs to compute the mean file lifetime across all measured lifetimes. Figure \ref{fig:lifetimes} shows the results of running this procedure once for each month in the span of January 2021-September 2021. The mean file lifetime across all months in this span is $0.968$ days. Figure \ref{fig:lifetime_hist_hours} shows a histogram summarizing the distribution of file lifetimes across the same time range. Figure \ref{fig:medium_lifetimes} shows a closer view of a subset of the lifetime distribution. This subset of the distribution roughly follows a power-law distribution, so Figure \ref{fig:medium_lifetimes} also includes the plot of a power law function fitted to the curve. This equation can be seen in Eq. \ref{eq:power_law}, and the values of its parameters can be seen in Table \ref{tab:my_label}.  

\begin{table}[]
    \centering
    \begin{tabular}{|c|c|c|}
         \hline
    < 1 Hour & < 5 Hours & < 10 Hours  \\
        \hline
       54.6\%  & 78\% & 83.8\% \\
       \hline
    \end{tabular}
    \vspace*{5mm}
    \caption{Percentages of file lifetimes falling under certain thresholds}
    \label{tab:read_percentages}
\end{table}

\begin{equation}
    f(x) = ax^{b} + \epsilon
    \label{eq:power_law}
\end{equation}

\begin{table}[]
    \centering
    \begin{tabular}{|c|c|c|}
        \hline
        $a$ & $b$ & $\epsilon$ \\
        \hline
        15227.387  & -1.031 & -995.488 \\
        \hline
    \end{tabular}
    \vspace*{5mm}
    \caption{Parameter values for Eq. \ref{eq:power_law} plotted in Figure \ref{fig:medium_lifetimes}}
    \label{tab:my_label}
\end{table}

\subsection{Data Transfer Size}
\label{subsec:data_sz}
\begin{figure}
    \centering
    \includegraphics[width=0.9\linewidth]{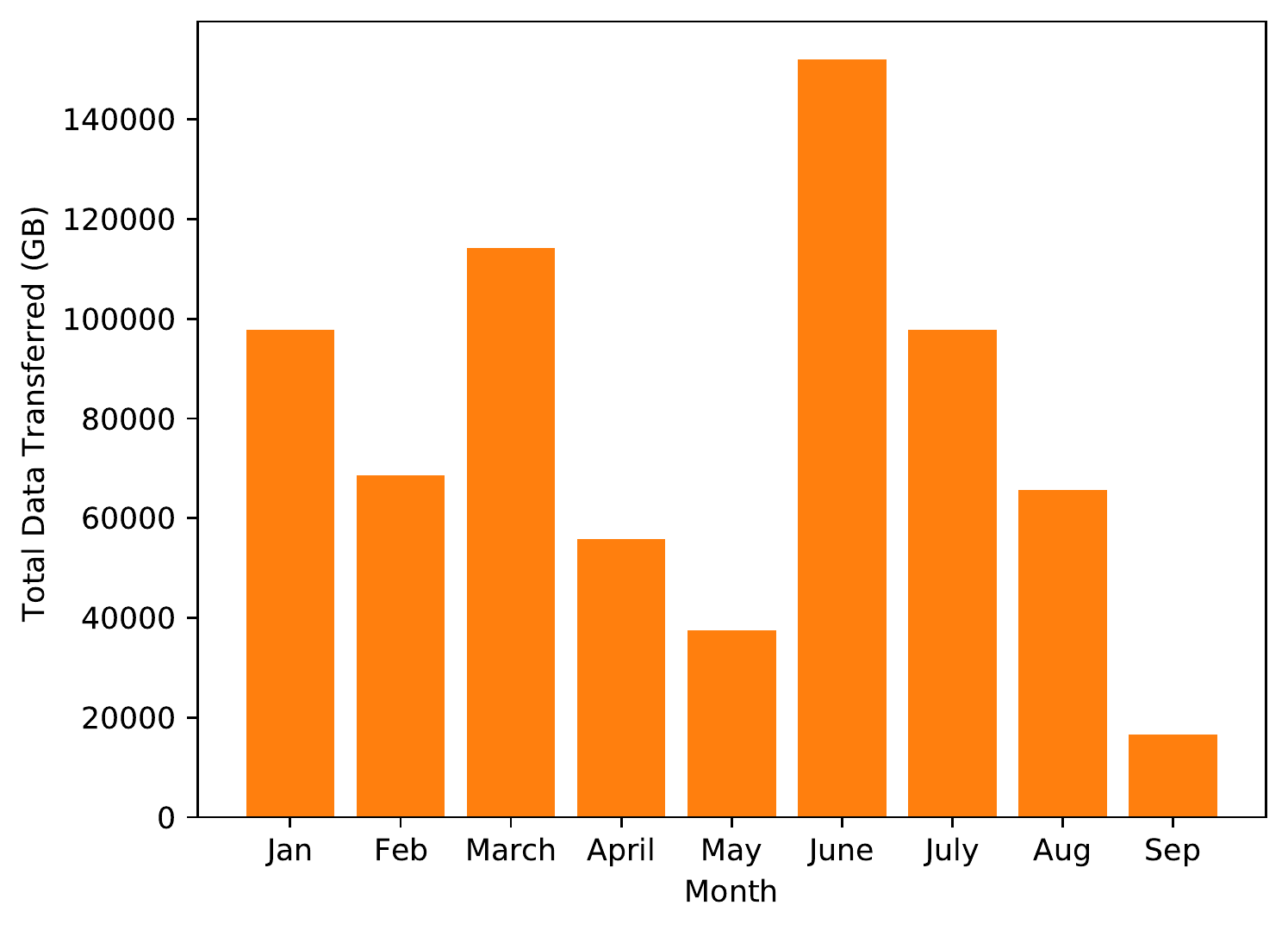}
    \caption{Total amount of data transferred to the cache for Jan 2021-Sep 2021}
    \label{fig:byte_totals}
\end{figure}

File transfers to the cache are denoted by the keyphrase \\ \texttt{"successfuly read size from info file = NNNNN"}, where \texttt{NNNNN} is an integer that denotes the byte-size of the file being transferred to the cache\footnote{The misspelling of the world "successfuly" is intentional, and accurately reflects the contents of the server logs}. By parsing XRootD server logs while searching for this keyphrase, it is possible to compute the total amount of data transferred to the cache during a given time frame. Our procedure parses each line in a set of XRootD logs while searching for the keyphrase. Upon finding a line with the keyphrase, it extracts the size of the file from the line and adds it to a sum $s$. At the end of the procedure, $s$ is the total amount of data transferred to the cache within the specified time frame of analysis. The results of running this procedure for a timeframe of Jan 2021-Sep 2021 are summarized in Figure \ref{fig:byte_totals}.

\section{Distributed Cache Node Simulation}
\label{sec:cachesim}

The procedures described in Section \ref{sec:lifecycle} opened up the possibility of developing a tool with the ability to simulate the behavior of an XRootD cache node on an access cycle corresponding to an arbitrary contiguous set of XRootD server logs. In order to simulate a cache node, we need to know what files are transferred to the cache, as well as how large these files are. The first requirement is satisfied by a procedure described in the following section. The second is covered by the procedure described in Section \ref{subsec:data_sz}. The ability to simulate cache behavior opens up a wide range of potential avenues for exploration. This section describes the various insights gained from this cache simulator. 

\subsection{Simulator Design and Implementation}
The behavior of an XRootD cache node is simulated by employing an LRU eviction policy. The core design element of the simulator is an ordered dictionary with relative paths to a file as keys and values as objects corresponding to a file instance. These file objects include the file size, file name, and the timestamp corresponding to the file's first access \footnote{The first access timestamp was used only for debugging purposes}.


To simulate cache behavior for a given access cycle, the start and end date of an access pattern need to be specified along with the cache's capacity.

From there, the software retrieves the XRootD logs corresponding to each day in the analysis timeframe. The keyphrase \\ \texttt{"successfuly read size from info file = NNNNN"} corresponds to a file transfer into the cache where \texttt{NNNNN} is an integer that denotes the byte-size of the file being transferred. Server log lines that include this keyphrase contain information regarding the total number of bytes transferred into the cache for a given file, as well as the file path\footnote{This information is elsewhere in a server log line that includes the keyphrase}. These lines denote cache misses. Upon encountering one of these lines, the simulator creates a new file object instance, and adds it to the front of the ordered dictionary, using the file path as the key. The simulator then checks to see if the total size of all the files in the ordered dictionary exceeds the cache size. If the cache size is exceeded, it evicts the last element in the ordered dictionary. When the simulator encounters read operations or readV operations, it uses a similar process as the one described in \ref{subsec:file_stats} to match these operations with file paths. If the file path is contained in the ordered dictionary, then the corresponding element is moved to the front of the ordered dictionary, and the operation is counted as a cache hit. Otherwise, the operation is counted as a cache miss. The total number of file read operations is also counted. This cache simulator has three modes of operation. Each are described in their respective subsections below.

\subsection{Hit Rate}
\begin{figure} 
    \centering
    \includegraphics[width=0.9\linewidth]{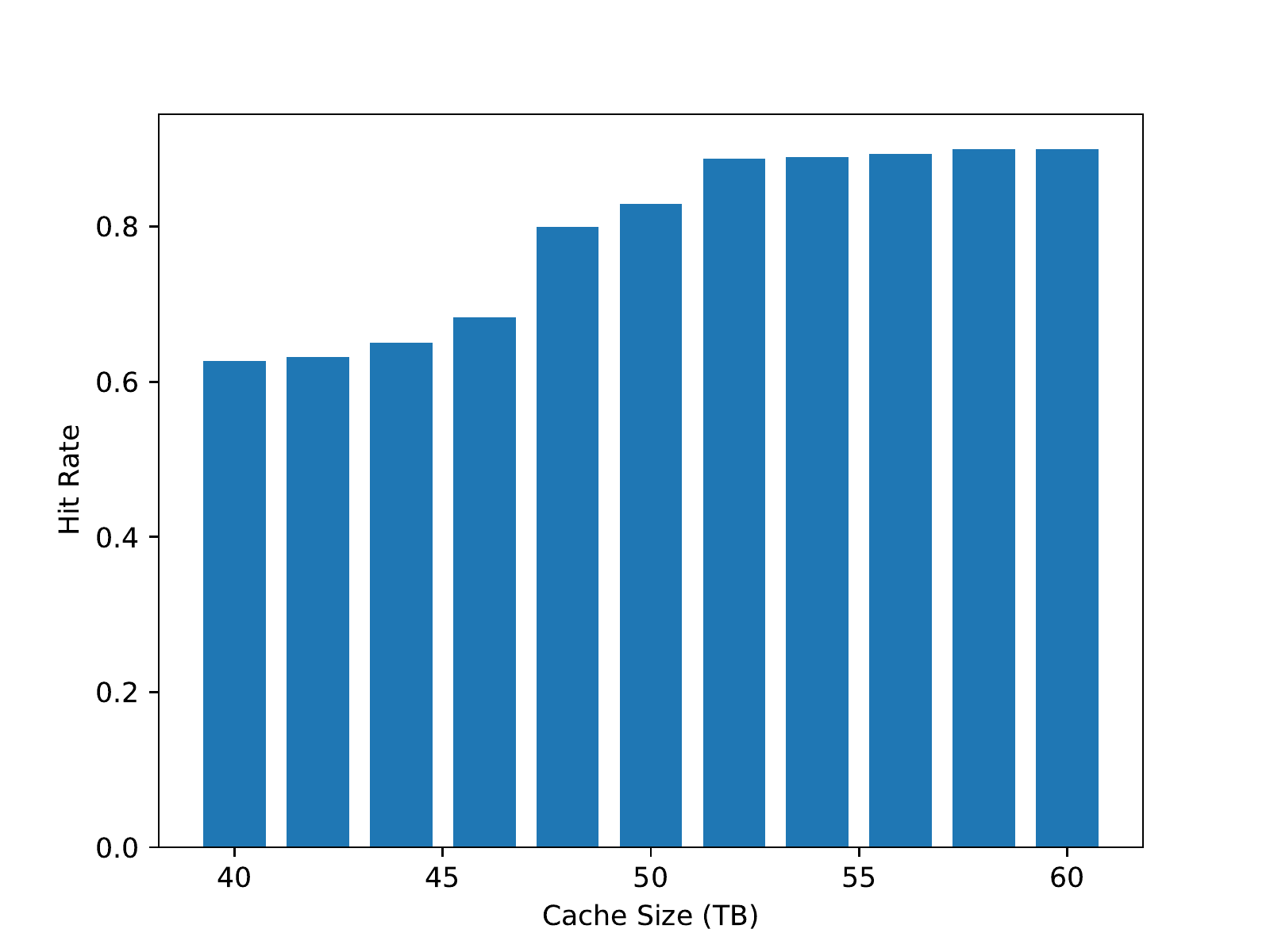}
    \caption{Simulated cache hit rates for a range of cache sizes (40TB-60TB) based on file access patterns of August 2021.  Note that as cache size varies from 40TB to 54TB, the cache hit rate goes from 0.62 (observed) to 0.89.}
    \label{fig:hit_rates}
\end{figure}

The first mode of operation calculates what the hit rate $h$ of a cache with a given capacity would be for a given access cycle.
The procedure described in the previous section is run, and the number of cache hits is divided by the total number of file read operations, returning the hit rate.
We ran the simulator in this mode on an access cycle spanning August 1st, 2021--August 31st, 2021.
The results of this run are summarized in Figure \ref{fig:hit_rates}. 
We observe that increasing the cache size from 40TB to about 56TB increase the cache hit rate from about 0.62 to 0.89.
This means the fraction of file accesses that require data transfer over wide-area network is reduced from 38\% to 15\%, a more than 2.5X reduction in the demand on the wide-area network.
Correspondingly, this would also reduce the data access time for the end users and improve the effectiveness of the overall storage cache system.



\subsection{Cache Content Modeling}
\begin{figure}
    \centering
    \includegraphics[width=0.9\linewidth]{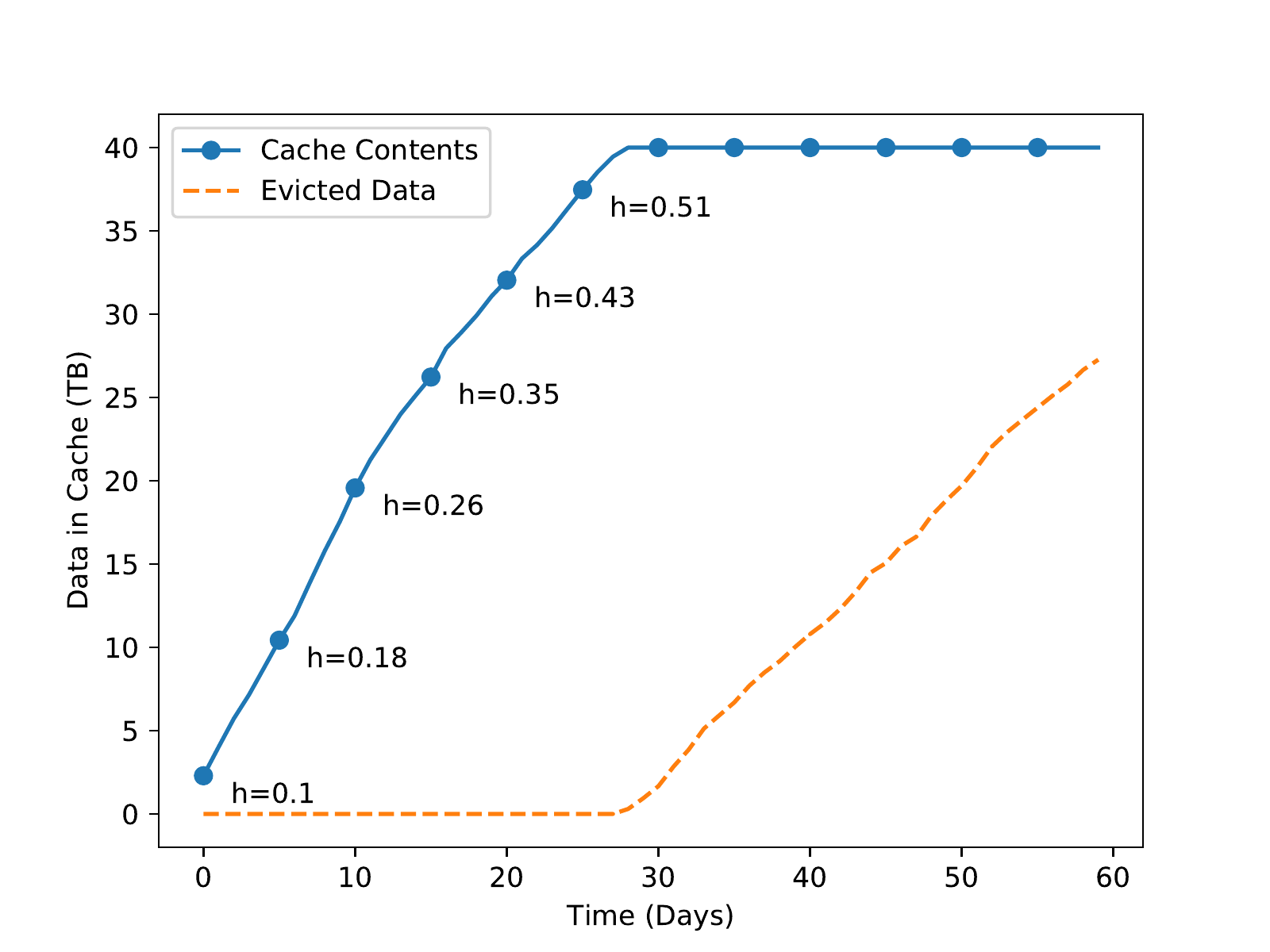}
    \caption{Cache hit rate as a function of time and varying hit rates (Cache size=40TB)}
    \label{fig:mode_2}
	\vspace{-0.7cm}
\end{figure}

The second mode of operation calculates the byte-size of a cache's contents and the total size of evicted data as a function of time $t$ and the hit rate $h$. The pseudo-code for this mode of operation can be seen in Algorithm 1. The values of the \texttt{"access\_rates"} and \texttt{"file\_size"} parameters were both computed using XRootD access data. The \texttt{"size\_params"} parameter and the \texttt{"rate\_params"} parameter are both arrays of random scalar values ranging from -2.0 to 2.0. The range of [-2.0, +2.0] was chosen to ensure sufficient deviation from the values of the \texttt{"access\_rates"} and \texttt{"file\_size"} parameters in both directions. At each time step, a random element $s_t$ is extracted from \texttt{size\_params}, and a random element $r_t$ is extracted from \texttt{rate\_params}. $s_t$ and $r_t$ are multiplied by \texttt{file\_size} and \texttt{access\_rates} respectively, and the products are used to compute the final cache size for the time step. The purpose of the random parameters is to introduce variances in the file size and file access rates, as this more closely models the real-world behavior. In addition, the \texttt{"access\_rates"} and \texttt{"file\_size"} parameters are both constant across timesteps, so variance is necessary to prevent the simulation from simply depicting linear growth. 
Once the cache is filled up, the simulator begins measuring the amount of data evicted from the cache. It should also be noted that at each time step, the hit rate $h$ increases by a small amount. 
We would expect a cache's hit rate to improve as the cache fills up, as more data in a cache means there's a higher chance of a cache hit. Thus, we increase $h$ as our simulated cache's contents grow in size. We ran the simulator in this mode for a span of 2 months. The results of this experiment are summarized in Figure \ref{fig:mode_2}.


\begin{algorithm}
    \label{alg:cache_sz}
    \caption{Cache Size as a Function of Hit Rate and Time}
    \begin{algorithmic}[1]
        \Procedure{modelCache}{$start\_date, end\_date$, $size\_params$, $rate\_params$}
            \State access\_rates $\gets 7000$
            \State file\_size $\gets 200,000,000$
            \State $t \gets start\_date-end\_date$
            \State $hit\_rate \gets 0.1 $
            \For{$i$ in $0:t$}
                \State $size\_param \gets size\_params[random\_index]$ 
                 \State $rate\_param \gets rate\_params[random\_index]$ 
                \State $val \gets (access\_rate*rate\_param)*(1-hit\_rate)*(file\_size*size\_param)$ 
                \If {$hit\_rate < 0.6$}
                \State $hit\_rate \gets hit\_rate+\delta$ \Comment{$\delta$ is a constant}
                \EndIf
                \State $result.append(val)$
            \EndFor
            
            \Return $result$
         \EndProcedure
    \end{algorithmic}
\end{algorithm}

\subsection{Cache Fill Up Times}
\begin{figure}
    \centering
    \includegraphics[width=0.9\linewidth]{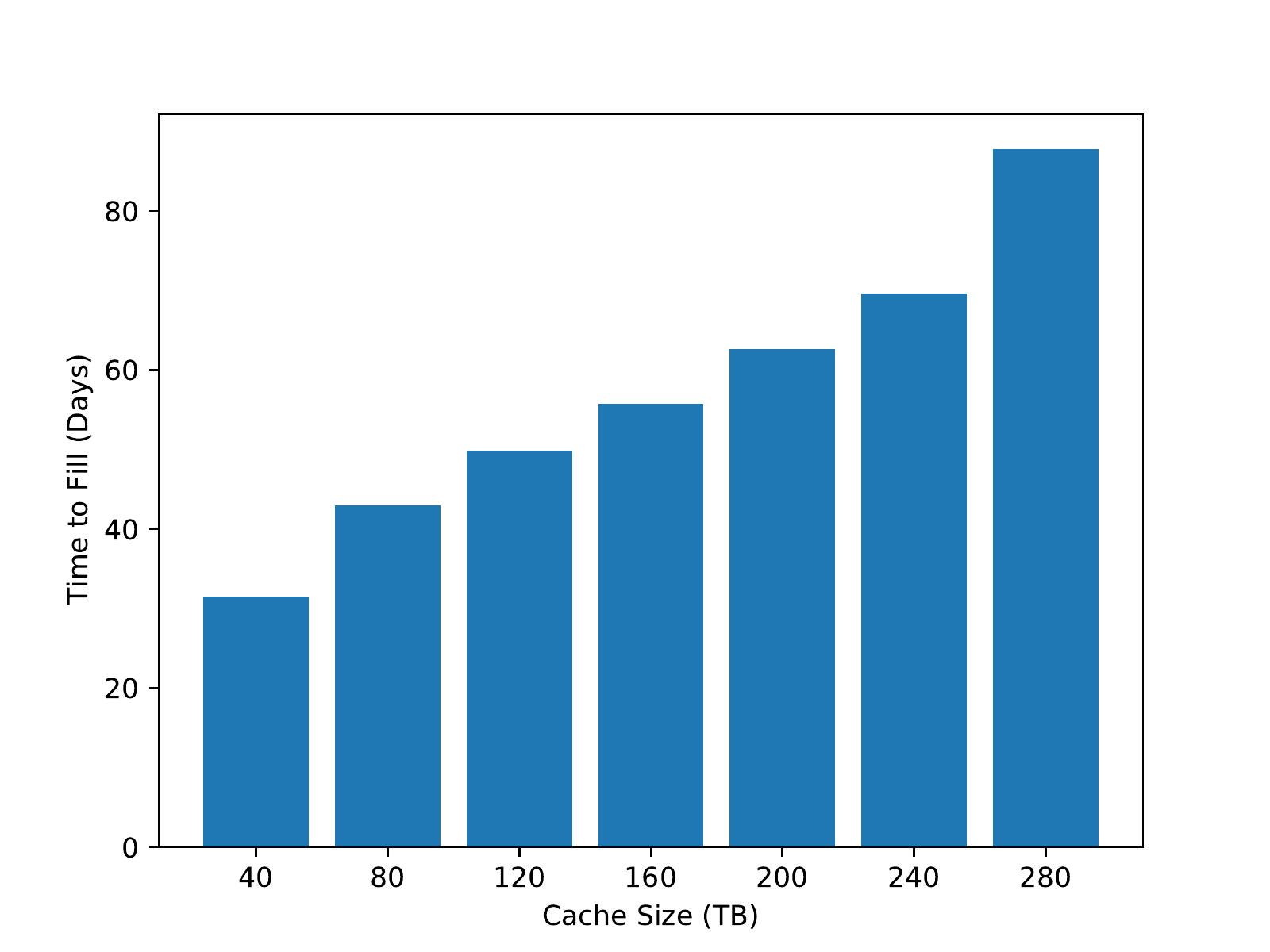}
    \caption{Time it takes to fill caches of sizes 40TB-280TB}
    \label{fig:cache_fillup}
	\vspace{-0.3cm}
\end{figure}

The third mode of operation calculates how long it takes to fill up caches of various sizes. The procedure described previously is run for a large access cycle. When the total contents of the cache equal to the cache size, the simulation is terminated, and the time stamp of the final file transfer is recorded. This time stamp is compared with the beginning time stamp of the access cycle to compute the total amount of time it took to fill the cache. We ran this mode of operation on a range of cache sizes from 40TB to 280TB, processing an access cycle spanning May 1st, 2021--August 31st, 2021. The results are summarized in Figure \ref{fig:cache_fillup}.

\section{Discussion}
\label{sec:discussion}

\subsection{File Access Patterns}
Figure \ref{fig:read_rates} demonstrates that the mean number of file read operations issued towards a single file varies substantially from month-to-month. The mean peaks in March 2021 at around 3500 read operations, and the mean is at its lowest in February 2021, at approximately 800 read operations. 
Figure \ref{fig:big_reads} demonstrates that despite the relatively large global mean of 1562.46, the bulk of files are issued fewer than 500 read operations in a given month. Figure \ref{fig:reads_fine_grained} further demonstrates that the majority of files are issued fewer than 200 read operations throughout the course of a month. Note that Figure \ref{fig:reads_fine_grained} roughly follows a bimodal distribution, with peaks at 25 and 150 read operations.

Figures \ref{fig:tot_reads}, \ref{fig:mean_reads}, and \ref{fig:offsets} provide more information regarding file read operations. Figure \ref{fig:tot_reads} shows that the total amount of data read from files differs greatly from month-to-month. June has the highest total, at approximately 139 Terabytes, while May has the smallest total, coming in at around 36.5 Terabytes. There is also a minor positive trend with respect to the total. Figure \ref{fig:mean_reads} demonstrates that there is a strong positive trend from month-to-month with respect to the mean size of a file read operation. This indicates that as the year progresses, read operations become larger. Figure \ref{fig:offsets} has the smallest spread of any figure in this paper. There is very little change in the mean file offset size from month to month, which indicates that read offsets tend to be consistent. 

From Figure \ref{fig:lifetimes}, the mean file lifetime is $0.968$ days, or 23.23 hours. However, Figure \ref{fig:lifetime_hist_hours} demonstrates that the distribution of file lifetimes is right-skewed, so this mean is inflated by the small number of large lifetimes. Figure \ref{fig:medium_lifetimes} and Table \ref{tab:read_percentages} further support this idea, as they show that the majority of file lifetimes are less than 10 hours, despite the much higher global mean. Therefore, if using file lifetimes to inform caching policy, it would be best to look to the distribution of lifetimes instead of the mean. 


\subsection{Cache Simulation}
Figure \ref{fig:hit_rates} shows a clear pattern with respect to cache hit rates as a function of the cache size. The hit rate $h$ starts at $\approx 0.626$ for a cache size of 40TB. $h$ gradually scales with the cache size, with a slightly larger-than-normal jump between 48TB and 50TB. The rate at which $h$ increases begins to flatten after 54TB. This indicates that a cache size of 52TB eliminates the majority of capacity misses, leaving primarily compulsory and conflict misses. Figure \ref{fig:byte_totals} further supports this argument, as it demonstrates that in August 2021 (the month the simulation was run for), approximately 60 TB were transferred into the cache due to cache misses, which is close to the cache size beyond which we no longer see hit rate increases.

The observed cache hit rate for a 40TB XCache node in August 2021 is 59.3. This is lower than the hit rate produced by the simulation for the same cache size. The difference can be explained by two factors. First, the simulator employs an LRU eviction policy, which does not necessarily reflect the behavior of the real XCache. Second, the simulator assumes that the XCache is fully associative. This assumption in particular could significantly contribute to the larger hit rate, as fully associative caches tend to have higher hit rates than n-way associative caches.

Figure \ref{fig:cache_fillup} also shows a clear pattern. As one would expect, the time it takes to fill up a cache scales with respect to the cache size. From cache sizes of 40TB-240TB, the increase in fill-up time between cache sizes is more or less consistent, but from 240TB-280TB, the increase is much more than anything else observed in Figure \ref{fig:cache_fillup}.

Figure \ref{fig:mode_2} shows what would be expected. As time goes on, the amount of data in the cache gradually increases, until eventually the cache fills up entirely. At this point, the amount of data evicted from the cache begins to grow at an essentially consistent rate. Additionally, as the hit rate increases, the total amount of data in the cache increases less between time steps. This is expected, as a higher hit rate would mean that fewer data accesses bring new data into the cache. Also of note, the cache fills up after approximately 30 days, which is the same amount of time Figure \ref{fig:cache_fillup} indicates is necessary to fill up a cache of 40TB.


\section{Summary \& Next Steps}
\label{sec:summary}
To inform the design choices of XRootD caches, we studied the operational logs to understand the cache usage patterns.
This paper provides insights into file read operations, file lifetimes, and how various changes to a cache node affect its behavior.
We find that increasing the XRootD cache size improves the cache hit rate, yielding faster overall file access.
Additionally, increasing the cache size nearly linearly increases the time to fill the cache.

This work could be expanded upon in a number different ways.
First, the cache simulator described in Section \ref{sec:cachesim} could be expanded to model different eviction policies,
while the current simulator is limited to the LRU eviction policy.
Additionally, future work will attempt to refine the cache simulator so that it is able to simulate hit rates that would be achieved by different file read rates. 
This work could also be expanded upon by developing machine learning models that can predict when a certain file is likely to be evicted from the cache.
Such models have been shown to be effective in creating more precise and detailed file lifetime data \cite{luis2021}.
A more effective model could inform design choices for better distributed storage caches.

\begin{acks}
This work was supported by the Office of Advanced Scientific Computing Research, Office of Science, of the U.S. Department of Energy under Contract No. DE-AC02-05CH11231, and used resources of the National Energy Research Scientific Computing Center (NERSC). 
This work was also supported by the National Science Foundation through the grants OAC-2030508, OAC-1836650, MPS-1148698, PHY-1120138, and OAC-1541349.
\end{acks}

\bibliographystyle{ACM-Reference-Format}
\bibliography{main}

\end{document}